\documentclass[preprint,aps,floats,nofootinbib]{revtex4}

\input epsf.tex
\def\DESepsf(#1 width #2){\epsfxsize=#2 \epsfbox{#1}}
\def\etal{ {\em et al.}}

\newcommand{\no}{\nonumber\\}

\newcommand{\be}{\begin{equation}}
\newcommand{\ee}{\end{equation}}
\newcommand{\bea}{\begin{eqnarray}}
\newcommand{\eea}{\end{eqnarray}}
\newcommand{\Cdot}{\hspace{-1mm}\cdot\hspace{-1mm}}

\def\thebibliography#1{\centerline{\bf REFERENCES}
  \list{[\arabic{enumi}]}{\settowidth\labelwidth{[#1]}\leftmargin
  \labelwidth\advance\leftmargin\labelsep\usecounter{enumi}}
\def\newblock{\hskip .11em plus .33em minus -.07em}\sloppy
  \clubpenalty4000\widowpenalty4000\sfcode`\.=1000\relax}

\begin{document}
\preprint{\vbox{\hbox{}\hbox{}\hbox{KEK-TH-865}}}
\draft

\vspace*{0.5cm}

\title{Heavy Baryonic Decays of $\Lambda_b \to \Lambda \eta^{(\prime)}$ \\
and Nonspectator Contribution}

\author{ \vspace{0.5cm}
M.~R.~Ahmady$^a$\footnote{mahmady@mta.ca},~~
C.~S.~Kim$^b$\footnote{cskim@yonsei.ac.kr,~~
http://phya.yonsei.ac.kr/\~{}cskim/},~~
Sechul~Oh$^c$\footnote{scoh@post.kek.jp}~~ and~~
Chaehyun~Yu$^b$\footnote{chyu@cskim.yonsei.ac.kr}}

\address{ \vspace{0.3cm}
$^a$Department of Physics, Mount Allison University, Sackville, NB
E4L 1E6, Canada \\
$^b$Department of Physics and IPAP, Yonsei University, Seoul
120-479, Korea \\
$^c$Theory Group, KEK, Tsukuba, Ibaraki 305-0801, Japan
\vspace{1cm}}

\vspace*{0.5cm}

\begin{abstract}

\noindent We calculate the branching ratios of the hadronic $\Lambda_b$
decays to $\eta$ and $\eta'$ in the factorization approximation
where the form factors are estimated via QCD sum rules and the
pole model.  Our results indicate that, contrary to $B\to
K\eta^{(\prime )}$ decays, the branching ratios for
$\Lambda_b\to\Lambda\eta$ and $\Lambda_b\to\Lambda\eta'$ are more
or less the same in the hadronic $\Lambda_b$ transitions.  We
estimate the branching ratio of $\Lambda_b\to\Lambda\eta^{(\prime
)}$ to be $10.80 (10.32)\times 10^{-6}$ in QCD sum rules, and
$2.78 (2.96)\times 10^{-6}$ in the pole model.  We
also estimate the nonfactorizable gluon fusion contribution to
$\Lambda_b\to\Lambda\eta'$ decay by dividing this process into
strong and weak vertices.  Our results point to an enhancement of
more than an order of magnitude due to this mechanism.
\\
PACS number(s): 13.30.Eg, 14.20.Mr, 14.40.Aq
\end{abstract}
\maketitle

%%%%%%%%%%%%%%%%%%%%%%%%%%%%%%%%%%%%%%%%%%%%%%%%%%%%%%
%%%%%%%%%%%%%%%%%%%%%%%%%%%%%%%%%%%%%%%%%%%%%%%%%%%%%%
\newpage
\section{Introduction}

For the  last few years, different experimental groups have been
accumulating plenty of  data for the charmless hadronic $B$ decay
modes. CLEO, Belle and BaBar Collaborations are providing us with
the information on the branching ratio (BR) and the CP asymmetry
for different decay modes.   A clear picture is about to emerge
from these information.
Among the $B \to PP$ ($P$ denotes a pseudoscalar meson) decay
modes, the BR for  the decay $B^+ \to K^+ \eta'$  is
found to be larger than that expected  within the standard model
(SM). The observed BR for this  mode in three different
experiments are \cite{cleo,belle,babar}
\begin{eqnarray}
{\mathcal B}( B^{\pm} \to K^{\pm} \eta')  &=&
( 80^{+10}_{-9} \pm 7 ) \times 10^{-6}  ~~[{\rm CLEO}], \nonumber \\
\mbox{} &=& ( 77.9^{+6.2+9.3}_{-5.9-8.7}) \times 10^{-6}  ~~[{\rm
Belle}],  \nonumber \\
\mbox{} &=& ( 67 \pm 5 \pm 5 ) \times 10^{-6}  ~~[{\rm BaBar}].
\end{eqnarray}

In order to explain the unexpectedly large branching ratio for $B
\to K \eta'$, different assumptions have been proposed, e.g.,
large form factors \cite{kagan}, the QCD anomaly effect
\cite{atwood,ali1}, high charm content in $\eta^{\prime}$
\cite{halperin,ali2,cheng1}, a new mechanism in the Standard Model
\cite{du,ahmady1}, the perturbative  QCD approach \cite{kou}, the
QCD improved factorization approach \cite{beneke, koy}, or new physics
like supersymmetry without R-parity \cite{choudhury,dko1,dko2}.
Even though some of these approaches turn out to be
unsatisfactory, the other approaches are still waiting for being
tested by experiment. Therefore, it would be much more desirable if
besides using $B$ meson system, one can have an alternative way to
test the proposed approaches in experiment.

Weak decays of the bottom baryon $\Lambda_b$ can provide a fertile
testing ground for the SM.  $\Lambda_b$ decays
can also be used as an alternative and complimentary source of
data to $B$ decays, because the underlying quark level processes
are similar in both $\Lambda_b$ and $B$ decays.  For example,
$\Lambda_b \to \Lambda \eta^{(\prime)}$ decay involves similar
quark level processes as $B \to K \eta^{(\prime)}$, i.e.,  $b \to
q \bar q s$ ($q = u, ~d, ~s$). In the coming years, large number
of $\Lambda_b$ baryons are expected to be produced in hadron
machines, like Tevatron and LHC, and a high-luminosity linear
collider running at the $Z$ resonance. For instance, the BTeV
experiment, with a luminosity $2 \times 10^{32}$ cm$^{-2}$
s$^{-1}$, is expected to produce $2 \times 10^{11}$ $b \bar b$
hadrons per $10^7$ seconds \cite{stone}, which would result in the
production of $2 \times 10^{10}$ $\Lambda_b$ baryons per year of
running \cite{giri}.  One of peculiar properties of $\Lambda_b$
decays is that, unlike $B$ decays, these decays can provide
valuable information about the polarization of the $b$ quark.
Experimentally the polarization of $\Lambda_b$ has been measured
\cite{aod}.

In this work, we study $\Lambda_b \to \Lambda \eta^{(\prime)}$
decay.  Our goal is two-fold: (i) The calculation of the BR for
$\Lambda_b \to \Lambda \eta^{(\prime)}$ involves hadronic form
factors which are highly model-dependent.  Using different models
for the form factors, we calculate the BR for $\Lambda_b \to
\Lambda \eta^{(\prime)}$ and investigate the model-dependence of
the theoretical prediction. (ii) As an alternative test for a
possible mechanism explaining the large BR for $B^+ \to K^+
\eta'$, we examine the same mechanism using $\Lambda_b \to \Lambda
\eta'$ decay.  Among the mechanisms proposed for understanding the
large ${\cal B} (B^+ \to K^+ \eta')$, we focus on a nonspectator
mechanism presented in Refs. \cite{du,ahmady1}.  In this mechanism,
$\eta'$ is produced via the fusion of two gluons: one from the QCD
penguin diagram $b \to s g^*$ and the other one emitted by the
light quark inside the $B$ meson.  We calculate this nonspectator
contribution to the BR for $\Lambda_b \to \Lambda \eta'$ in order
to examine its validity. If this nonspectator process is indeed
the true mechanism responsible for the large ${\cal B} (B^+ \to
K^+ \eta')$, then the same mechanism would affect ${\cal B}
(\Lambda_b \to \Lambda \eta')$ as well. Thus, one can test the
validity of this mechanism in the future experiments such as BTeV,
LHC-b, etc., by comparing ${\cal B} (\Lambda_b \to \Lambda \eta')$
calculated with/without the nonspectator contribution with the
measured results.

We organize our work as follows.  In Sec. II, we present the
effective Hamiltonian for the usual $\Delta B =1 $ transition and
for the nonspectator process.  We calculate the BR for $\Lambda_b
\to \Lambda \eta^{(\prime)}$ decay without considering the
nonspectator mechanism in Sec. III.  The nonspectator
contribution to $\Lambda_b \to \Lambda \eta^{\prime}$ is estimated
in Sec. IV. We conclude in Sec. V.

%%%%%%%%%%%%%%%%%%%%%%%%%%%%%%%%%%%%%%%%%%%%%%%%%%%%%%%%%%
%%%%%%%%%%%%%%%%%%%%%%%%%%%%%%%%%%%%%%%%%%%%%%%%%%%%%%%%%%

\section{Effective Hamiltonian for $\Lambda_b \to \Lambda \eta^{(\prime)}$ decays}

The effective Hamiltonian $H_{\rm eff}$ for the $\Delta B =1 $
transition is
\begin{eqnarray}
H_{\rm eff} &=& \frac{4 G_F}{\sqrt{2}} \bigg[ V_{ub} V^\ast_{uq} (
c_1 O_{1u}^q + c_2 O_{2u}^q)
     + V_{cb} V^\ast_{cq} ( c_1 O_{1c}^q + c_2 O_{2c}^q) \nonumber \\
&& \hspace{1cm}    - V_{tb} V^\ast_{tq} \sum_{i=3}^{12} c_i O_i^q
\bigg] + h.c., \label{effham}
\end{eqnarray}
where $q=d ~{\rm or}~ s$, and
\begin{eqnarray}
&\mbox{}& O_{1f}^q = \bar{q}_\alpha \gamma_\mu L f_\alpha
        \bar{f}_\beta  \gamma^\mu L b_\beta, ~~
O_{2f}^q = \bar{q}_\alpha \gamma_\mu L f_\beta
        \bar{f}_\beta  \gamma^\mu L b_\alpha, \no
&\mbox{}& O_{3(5)}^q = \bar{q}_\alpha \gamma_\mu L b_\alpha
\sum_{q^\prime}^{} \bar{q^\prime}_\beta  \gamma^\mu L(R)
q^\prime_\beta, ~~ O_{4(6)}^q = \bar{q}_\alpha \gamma_\mu L
b_\beta \sum_{q^\prime}^{} \bar{q^\prime}_\beta \gamma^\mu L(R)
q^\prime_\alpha, \no &\mbox{}& O_{7(9)}^q =
\frac{3}{2}\bar{q}_\alpha \gamma_\mu L b_\alpha \sum_{q^\prime}^{}
e_{q^\prime} \bar{q^\prime}_\beta  \gamma^\mu R(L) q^\prime_\beta,
~~ O_{8(10)}^q = \frac{3}{2}\bar{q}_\alpha \gamma_\mu L b_\beta
\sum_{q^\prime}^{} e_{q^\prime} \bar{q^\prime}_\beta  \gamma^\mu
R(L) q^\prime_\alpha, \no &\mbox{}& O_{11} = \frac{g_s}{32 \pi^2}
m_b \bar{q} \sigma^{\mu\nu} R T^a b G^a_{\mu\nu}, ~~ O_{12} =
\frac{e}{32 \pi^2} m_b \bar{q} \sigma^{\mu\nu} R b F_{\mu\nu},
\end{eqnarray}
with $f=u ~{\rm or}~c$ and $q^\prime = u, d, s, c$ and $L(R) =
(1\mp\gamma_5)/2$. The SU(3) generator $T^a$ is normalized as
${\rm Tr}(T^a T^b) = \frac{1}{2} \delta^{ab}$. $\alpha$ and
$\beta$ are  the color indices.  $G^{\mu \nu}_a$ and $F^{\mu \nu}$
are the gluon and photon field strength, and $c_i$'s are the
Wilson coefficients (WCs). We use the improved effective WCs given
in Refs. \cite{chen,hou}.  The renormalization scale is taken to be
$\mu=m_b$ \cite{ddo}. The operators $O_1$, $O_2$ are the tree
level and QCD corrected operators, $O_{3-6}$ are the gluon induced
strong penguin operators, and finally  $O_{7-10}$ are the
electroweak penguin operators due to $\gamma$ and $Z$ exchange,
and  the box diagrams at loop level. In this work we shall take
into account the chromomagnetic operator $O_{11}$, but neglect the
extremely small contribution from $O_{12}$.

Considering the gluon splits into two quarks, the chomomagnetic
operator is rewritten in the Fierz transformed form as
\begin{eqnarray}
O_{11} = \frac{\alpha_s}{16 \pi } \frac{m_b^2}{k^2} \frac{N_c^2-1}{N_c^2}
\left[ \delta_{\alpha\beta} \delta_{\alpha^\prime \beta^\prime} -
\frac{2N_c}{N_c^2-1} T_{\alpha\beta}^a T_{\alpha^\prime
\beta^\prime}^a \right] \sum_{i=1}^{4} T_i,
\end{eqnarray}
where
\begin{eqnarray}
T_1&=& 2 \bar{s}_\alpha \gamma^\mu L q^\prime_\beta
\bar{q^\prime}_{\alpha^\prime} \gamma_\mu L b_{\beta^\prime} -4
\bar{s}_\alpha R q^\prime_\beta \bar{q^\prime}_{\alpha^\prime} L
b_{\beta^\prime}, \no T_2&=& \frac{m_s}{m_b}(2 \bar{s}_\alpha
\gamma^\mu R q^\prime_\beta
    \bar{q^\prime}_{\alpha^\prime} \gamma_\mu R b_{\beta^\prime}
-4 \bar{s}_\alpha L q^\prime_\beta \bar{q^\prime}_{\alpha^\prime}
R b_{\beta^\prime}, \no T_3&=& \frac{(p_b+p_s)_\mu}{m_b} [
\bar{s}_\alpha \gamma^\mu L q^\prime_\beta
\bar{q^\prime}_{\alpha^\prime} R b_{\beta^\prime} + \bar{s}_\alpha
R q^\prime_\beta \bar{q^\prime}_{\alpha^\prime} \gamma^\mu R
 b_{\beta^\prime}], \no
T_4&=& i\frac{(p_b+p_s)_\mu}{m_b} [ \bar{s}_\alpha \sigma^{\mu\nu}
R q^\prime_\beta \bar{q^\prime}_{\alpha^\prime} \gamma_\nu R
b_{\beta^\prime} - \bar{s}_\alpha \gamma_\nu L q^\prime_\beta
\bar{q^\prime}_{\alpha^\prime} \sigma^{\mu \nu} R
b_{\beta^\prime}].
\end{eqnarray}
Here $p_b$ and $p_s$ are the four-momenta of $b$- and $s$-quarks,
respectively.  $N_c$ denotes the effective number of colors and $k
\equiv p_b - p_s$ is the gluon momentum. In the heavy quark limit,
$k^2 = m_b^2 (1-x)$, where $x$ is the momentum fraction of
$\eta^{(\prime)}$. The average gluon momentum can be estimated
\cite{bensalem} as \be \left\langle \frac{m_b^2}{k^2}
\right\rangle = \int_0^1 (\phi_{\eta^{(\prime)}}(x) m_b^2 / k^2)
dx, \ee where $\phi_{\eta^{(\prime)}}$ is the $\eta^{(\prime)}$
light-cone distribution and its asymptotic form is
$\phi_{\eta^{(\prime)}}=6x(1-x)$.

The effective Hamiltonian for the nonspectator contribution can be
obtained by considering the dominant chromo-electric component of
the QCD penguin diagram \cite{ahmady1,ahmady2}:
\begin{eqnarray}
H_{\rm nonsp} = i C H [ \bar s \gamma^{\mu} (1-\gamma_5) T^a b ] (
\bar q \gamma^{\nu} T^a q ) {1 \over p_2^2} \epsilon_{\mu \nu \rho
\sigma} p_1^{\rho} p_2^{\sigma}~, \label{hnonsp}
\end{eqnarray}
where
\begin{eqnarray}
C = {G_F \over \sqrt{2}} {\alpha_s \over 2 \pi} V_{tb} V^*_{ts} [
E(x_t) -E(x_c) ]~,
\end{eqnarray}
$q$ denotes the spectator quark, and $p_i$ $(i=1,2)$ are the
four-momenta of the two gluons relevant to the $g -g -\eta'$
vertex. The coefficient function $E$ is defined as
\begin{eqnarray}
E(x_i) = - {2 \over 3} {\rm ln} x_i + {x^2_i (15 -16 x_i +4 x_i^2)
\over 6 (1 -x_i)^4} {\rm ln} x_i + {x_i (18 -11 x_i -x_i^2) \over
12 (1 -x_i)^3} ~,
\end{eqnarray}
where $x_i = m_i^2 / m_W^2$ with
$m_i$ being the internal quark mass. $H$ is the form factor
parametrizing the $g -g -\eta'$ vertex
\begin{eqnarray}
A_{\mu \nu} (g g \to \eta') = i H(p_1^2, p_2^2, m_{\eta'}^2 )
\delta^{ab} \epsilon_{\mu \nu \rho \sigma} p_1^{\rho} p_2^{\sigma}
~.
\end{eqnarray}
Using the decay mode $\psi \to \eta' \gamma$, $H (0,0,m_{\eta'}^2
)$ is estimated to be approximately 1.8 GeV$^{-1}$.

%%%%%%%%%%%%%%%%%%%%%%%%%%%%%%%%%%%%%%%%%%%%%%%%%%%%%%%%%%
%%%%%%%%%%%%%%%%%%%%%%%%%%%%%%%%%%%%%%%%%%%%%%%%%%%%%%%%%%

\section{$\Lambda_{b} \to \Lambda \eta^{(\prime)}$ Decay Process
within Factorization Approach}

In general, the vector and axial-vector matrix elements for the
$\Lambda_b \to \Lambda$ transition can be parameterized as
\begin{eqnarray}
\langle \Lambda | \bar{s} \gamma_\mu b | \Lambda_b \rangle &=&
\bar{u}_\Lambda \left[ f_1 \gamma_\mu +i \frac{f_2}{m_{\Lambda_b}}
\sigma_{\mu\nu} q^\nu +\frac{f_3}{m_{\Lambda_b}}q_\mu \right]
u_{\Lambda_b}, \no \langle \Lambda | \bar{s} \gamma_\mu \gamma_5 b
| \Lambda_b \rangle &=& \bar{u}_\Lambda \left[ g_1 \gamma_\mu
\gamma_5+i \frac{g_2}{m_{\Lambda_b}} \sigma_{\mu\nu}q^\nu\gamma_5
+\frac{g_3}{m_{\Lambda_b}}q_\mu\gamma_5 \right] u_{\Lambda_b},
\end{eqnarray}
where the momentum transfer $q^{\mu}
=p^{\mu}_{\Lambda_b}-p^{\mu}_\Lambda$ and $f_i$ and $g_i
(i=1,2,3)$ are Lorentz invariant form factors. Alternatively, with
the HQET, the hadronic matrix elements for the $\Lambda_b \to
\Lambda$ transition can be parameterized  \cite{mannel} as
\begin{eqnarray}
\langle \Lambda | \bar{s} \Gamma b | \Lambda_b \rangle =
\bar{u}_\Lambda \left[ F_1(q^2)  + v\hspace{-2mm}/
\hspace{0.3mm}F_2(q^2) \right] \Gamma  u_{\Lambda_b},
\end{eqnarray}
where $v=p_{\Lambda_b}/m_{\Lambda_b}$ is the four-velocity of
$\Lambda_b$ and $\Gamma$ denotes the possible Dirac matrix. The
relations between $f_i, g_i$ and $ F_i$ can be easily given by \be
f_1=g_1=F_1+r F_2,~~ f_2=f_3=g_2=g_3=F_2, \ee where
$r=m_{\Lambda}/m_{\Lambda_b}$.

The decay constants of the $\eta$ and $\eta'$ mesons,
$f^q_{\eta^{(\prime)}}$, are defined by
\begin{equation}
\langle 0 | \bar q \gamma^{\mu} \gamma_5 q | \eta^{(\prime)}
\rangle = i f^q_{\eta^{(\prime)}} p^{\mu}_{\eta^{(\prime)}} ~~~ (q
= u, s).
\end{equation}
Due to the $\eta - \eta'$ mixing, the decay constants of the
physical $\eta$ and $\eta'$ are related to those of the flavor
SU(3) singlet state $\eta_0$ and octet state $\eta_8$ through the
relations \cite{leut,feldmann}
\begin{eqnarray}
f^u_{\eta} &=& {f_8 \over \sqrt{6}} \cos \theta_8 - {f_0 \over
\sqrt{3}} \sin \theta_0 ~, ~~ f^s_{\eta} = -2 {f_8 \over \sqrt{6}}
\cos \theta_8 - {f_0 \over \sqrt{3}} \sin \theta_0 ~, \nonumber \\
f^u_{\eta'} &=& {f_8 \over \sqrt{6}} \sin \theta_8 + {f_0 \over
\sqrt{3}} \cos \theta_0 ~, ~~ f^s_{\eta'} = -2 {f_8 \over
\sqrt{6}} \sin \theta_8 + {f_0 \over \sqrt{3}} \cos \theta_0 ~,
\label{femix}
\end{eqnarray}
where $\theta_8$ and $\theta_0$ are the mixing angles and
phenomenologically $\theta_8 = -21.2^0$ and $\theta_0 = -9.2^0$
\cite{feldmann}.  We use $f_8 = 166$ MeV and $f_0 =154$ MeV \cite{chen}.

The decay amplitude of $\Lambda_b \to \Lambda \eta^{\prime}$ is
given  \cite{bensalem} by
\begin{eqnarray}
\langle \Lambda \eta^\prime | H_{\rm eff} | \Lambda_b \rangle &=&
\frac{4 G_F}{\sqrt{2}} \bigg[ V_{ub} V_{us}^\ast a_2 \langle
\eta^\prime | \bar{u} \gamma^\mu L u | 0 \rangle \langle \Lambda |
\bar{s} \gamma_\mu L b | \Lambda_b \rangle \no && \hspace{5mm} -
V_{tb} V_{ts}^\ast
\left(2a_3-2a_5-\frac{1}{2}a_7+\frac{1}{2}a_9\right) \langle
\eta^\prime | \bar{u} \gamma^\mu L u | 0 \rangle \langle \Lambda |
\bar{s} \gamma_\mu L b | \Lambda_b \rangle \no && \hspace{5mm} -
V_{tb} V_{ts}^\ast
\left\{a_3+a_4-a_5+\frac{1}{2}a_7-\frac{1}{2}a_9
+\left(1+\frac{2p_b \Cdot q}{m_b^2}\right)a_f\right\}\no
&&\hspace{2cm}\times \langle \eta^\prime | \bar{s} \gamma^\mu L s
| 0 \rangle \langle \Lambda | \bar{s} \gamma_\mu L b | \Lambda_b
\rangle \no && \hspace{5mm} - V_{tb} V_{ts}^\ast
\chi_{\eta^\prime} \left(-a_6+\frac{1}{2}a_8-\frac{5}{4}a_f\right)
\langle \eta^\prime | \bar{u} \gamma^\mu L u | 0 \rangle \langle
\Lambda | \bar{s} \gamma_\mu R b | \Lambda_b \rangle \no &&
\hspace{5mm} - V_{tb} V_{ts}^\ast  \chi_{\eta^\prime}
\left(a_6-\frac{1}{2}a_8+\frac{5}{4}a_f\right) \langle \eta^\prime
| \bar{s} \gamma^\mu L s | 0 \rangle \langle \Lambda | \bar{s}
\gamma_\mu R b | \Lambda_b \rangle \bigg],   \label{amp}
\end{eqnarray}
where
\begin{eqnarray}
a_i &\equiv& c_i^{\rm eff} + \frac{1}{N_c} c_{i+1}^{\rm eff}
~~~({\rm for}~ i = {\rm odd}), \no a_i &\equiv& c_i^{\rm eff} +
\frac{1}{N_c} c_{i-1}^{\rm eff} ~~~({\rm for}~ i = {\rm even}),
\no \chi_{\eta^\prime} &=& \frac{m_{\eta^\prime}^2}{m_b m_s}, \no
a_f &=& \frac{\alpha_s}{16 \pi k^2} m_b^2 \frac{N_c^2-1}{N_c^2}
c_{11}.
\end{eqnarray}
In the above amplitude, we have taken into account the anomaly
contribution \footnote{This anomaly contribution was not taken
into account in Ref. \cite{bensalem}.} to the matrix element
$\langle \eta' |\bar s \gamma_5 s |0 \rangle$
\cite{ali2,ddo,ball,cheng2}, which leads to
\begin{eqnarray}
\langle \eta' |\bar s \gamma_5 s | 0 \rangle = i {(f^s_{\eta'}
-f^u_{\eta'}) m^2_{\eta'} \over 2 m_s}~.
\end{eqnarray}
The similar expression for the decay amplitude of $\Lambda_b \to
\Lambda \eta$ can be obtained by replacing $\eta'$ by $\eta$ in
the above Eq. (\ref{amp}).

The decay amplitude given in Eq. (\ref{amp}) can be rewritten in
the general form
\begin{eqnarray}
{\cal M} \equiv \langle \Lambda \eta^\prime | H_{\rm eff} |
\Lambda_b \rangle = i \bar{u}_\Lambda ( a+
b\gamma_5)u_{\Lambda_b}.
\end{eqnarray}
The averaged square of the amplitude is
\begin{eqnarray}
\overline{|{\cal M}|^2} = 2( |a|^2 -|b|^2) m_\Lambda m_{\Lambda_b}
+ 2( |a|^2 +|b|^2) p_\Lambda \cdot p_{\Lambda_b},
\end{eqnarray}
where
\begin{eqnarray}
a&=& (X+Y) \left[ (m_{\Lambda_b}-m_\Lambda) f_1
       + \frac{m_{\eta^\prime}^2}{m_{\Lambda_b}} f_3\right], \no
b&=& (X-Y) \left[ (m_{\Lambda_b}-m_\Lambda) g_1
       - \frac{m_{\eta^\prime}^2}{m_{\Lambda_b}} g_3\right], \no
X&=& \frac{G_F}{\sqrt{2}}
    \bigg[ \Big\{ V_{ub} V_{us}^\ast a_2
 -V_{tb} V_{ts}^\ast \Big(2 a_3 -2a_5 -\frac{1}{2} a_7 +\frac{1}{2} a_9\Big)
   \Big\} f_{\eta^\prime}^u \no
&&\hspace{6.5mm}
 -V_{tb} V_{ts}^\ast \Big\{ a_3 +a_4 -a_5 +\frac{1}{2} a_7 -\frac{1}{2} a_9
  +\Big( 1+\frac{2 p_b \cdot q}{m_b^2}\Big) a_f\Big\}
    f_{\eta^\prime}^s \bigg],\no
Y&=& \frac{G_F}{\sqrt{2}}
 V_{tb} V_{ts}^\ast \chi_{\eta^\prime}
\Big( a_6  -\frac{1}{2} a_8 +\frac{5}{4} a_f\Big)
(f_{\eta^\prime}^u - f_{\eta^\prime}^s).
\end{eqnarray}
Then the decay width of $\Lambda_b \to \Lambda \eta'$ in the rest
frame of $\Lambda_b$ is given by
\begin{eqnarray}
\Gamma (\Lambda_b \to \Lambda \eta') = \frac{1}{16 \pi
m_{\Lambda_b}}
\lambda^{\frac{1}{2}}\bigg(1,\frac{m_\Lambda^2}{m_{\Lambda_b}^2},
\frac{m_{\eta^\prime}^2}{m_{\Lambda_b}^2}\bigg) \sum|{\cal M}|^2,
\end{eqnarray}
where
\begin{equation}
\lambda(1,a,b)= 1+a^2 +b^2 -2a -2b -2ab.
\end{equation}

For numerical calculations, we need specific values for the form
factors in the $\Lambda_b \to \Lambda$ transition which are
model-dependent. We use the values of the form factors from both
the QCD sum rule approach \cite{huang} and the pole model
\cite{mannel,geng}.  In the QCD sum rule approach, the form
factors $F_1$ and $F_2$ are given by
\begin{eqnarray}
F_1 &=& - {e^{2 \bar \Lambda /M +m_{\Lambda}^2 /T} \over 2
f_{\Lambda_b} f_{\Lambda}} \int^{\nu_c}_{0} d\nu \int^{2 \nu
z}_{0} ds \rho^1_{pert} e^{-s /T - \nu /M} - {1 \over 3} \langle
\bar q q \rangle^2\nonumber \\
&\mbox{}& - {1 \over 32 \pi^4} \langle \alpha_s G G \rangle
\int^{T /4}_{0} \left( 1 - {4 \beta \over T} \right) e^{-4 \beta
(1 -4 \beta /T) /M^2 -8 \beta z / (TM)} d\beta, \nonumber \\
F_2 &=& - {e^{2 \bar \Lambda /M +m_{\Lambda}^2 /T} \over 2
f_{\Lambda_b} f_{\Lambda}} \int^{\nu_c}_{0} d\nu \int^{2 \nu
z}_{0} ds \rho^2_{pert} e^{-s /T - \nu /M}  \nonumber \\
&\mbox{}& + {1 \over 8 \pi^4} \langle \alpha_s G G \rangle \int^{T
/4}_{0} \left( 1 - {4 \beta \over T} \right) {\beta \over M} e^{-4
\beta (1 -4 \beta /T) /M^2 -8 \beta z / (TM)} d\beta,
\end{eqnarray}
where
\begin{eqnarray}
\rho^1_{pert} &=& {1 \over 32 \pi^4 \sigma^3} \{ -2 z^3 \sigma^3
-[ -s +z (\nu + 2z)]^3 + 3 z^2 [-s +z (\nu +2 z)] \sigma^2 \},
\nonumber \\
\rho^2_{pert} &=& - {1 \over 64 \pi^4 \sigma^3} [ s -2 z^2 + z (-
\nu +\sigma) ]^2 [ \nu s +8 z^3 -4 z^2 (-2 \nu +\sigma) -2 z (-
\nu^2 +5 s +\nu \sigma) ], \nonumber \\
\sigma &=& \sqrt{ -4 s + (\nu +2 z)^2 } ~.
\end{eqnarray}
Here $z= \frac{p_\Lambda \cdot p_{\Lambda_b}}{m_{\Lambda_b}}
=\frac{m_{\Lambda_b}^2 + m_{\Lambda}^2 - q^2}{2 m_{\Lambda_b}}$
$(q^{\mu} = p_{\Lambda_b}^{\mu} -p_{\Lambda}^{\mu} )$ and the
Borel parameter $M = \frac{4 T}{m_b}$.  For the other relevant
conventions and notation, we refer to Ref. \cite{huang}.  In Figs.
1 and 2 we show the form factors $F_1$ and $F_2$ as a function of
the Borel parameter $M = {4 T \over m_b}$ for $\Lambda_b \to
\Lambda \eta^{(')}$, respectively.  In $\Lambda_b \to \Lambda
\eta^{(')}$, $F_1 = 0.510 (0.514)$ and $F_2 = -0.058 (-0.060)$ for
$M = 1.5$, $F_1 = 0.476 (0.481)$ and $F_2 = -0.084 (-0.088)$ for
$M = 1.7$, and $F_1 = 0.473 (0.479)$ and $F_2 = -0.117 (-0.122)$
for $M = 1.9$~.  The BRs of $\Lambda_b \to \Lambda \eta'$ and
$\Lambda_b \to \Lambda \eta$ versus $\xi \equiv {1 \over N_c}$ for
different values of the Borel parameter $M = {4 T \over m_b}$ are
shown in Fig. 3 and Fig. 4, respectively.
Our result shows
\begin{equation}
{\cal B} (\Lambda_b \to \Lambda \eta') = (5.0 - 14.5) \times 10^{-6} ~,
\end{equation}
and
\begin{equation}
{\cal B} (\Lambda_b \to \Lambda \eta) = (5.8 - 13.7) \times
10^{-6} ~.
 \end{equation}
For $\xi = 1 /3 ~ ({\rm i.e.,} ~N_c = 3)$ and $M = 1.7$
GeV, ${\cal B} (\Lambda_b \to \Lambda \eta') = 8.93 \times
10^{-6}$ and ${\cal B} (\Lambda_b \to \Lambda \eta) = 9.15 \times
10^{-6}$. We recall that in the case of $B \to K \eta'$ a small
value of $\xi$ ($\xi \leq 0.1$) is favored to fit the experimental
data on the BR in the framework of the generalized factorization
\cite{ali2,chen,ddo,cheng2}.  In the figures the shaded region
denotes the case of $\xi \leq 0.1$~, favored from the analysis of
$B \to K \eta'$. For $\xi = 0.1$, ${\cal B} (\Lambda_b \to \Lambda
\eta') = 11.17 \times 10^{-6}$ and ${\cal B}
(\Lambda_b \to \Lambda \eta) = 10.83 \times 10^{-6}$.

%%%%%%%%%%%%%%%%%%%%%%%%%%%%%%%%%%%%%%%%%%%
\begin{figure}
    \centerline{ \DESepsf(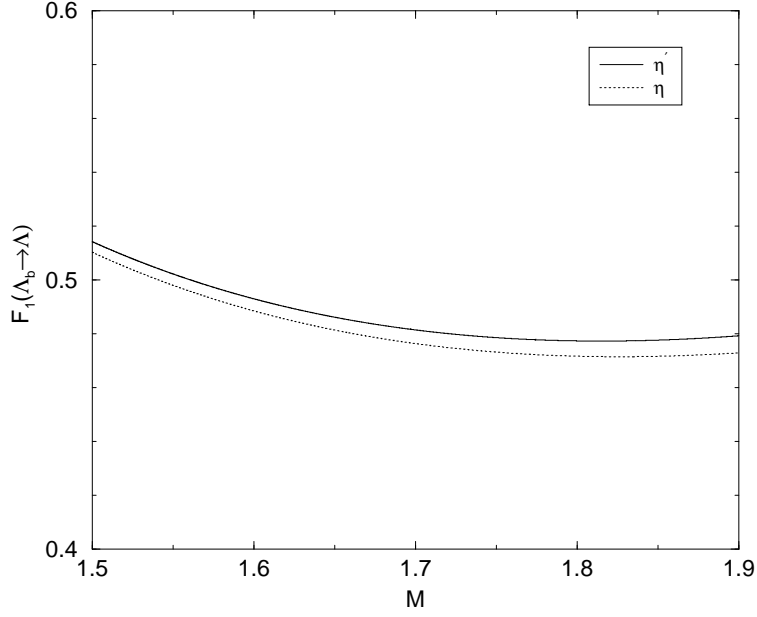 width 10cm)}
    \smallskip
    \caption{\label{fig:fig1} The form factor $F_1$ for the
    transition $\Lambda_b \to \Lambda$ versus the Borel parameter
    $M ~(= {4 T \over m_b})$.
    The dotted (solid) line corresponds to the case of
    $\Lambda_b \to \Lambda \eta^{(')}$.}
\end{figure}
%%%%%%%%%%%%%%%%%%%%%%%%%%%%%%%%%%%%%%%%%%%

%%%%%%%%%%%%%%%%%%%%%%%%%%%%%%%%%%%%%%%%%%%
\begin{figure}
    \centerline{ \DESepsf(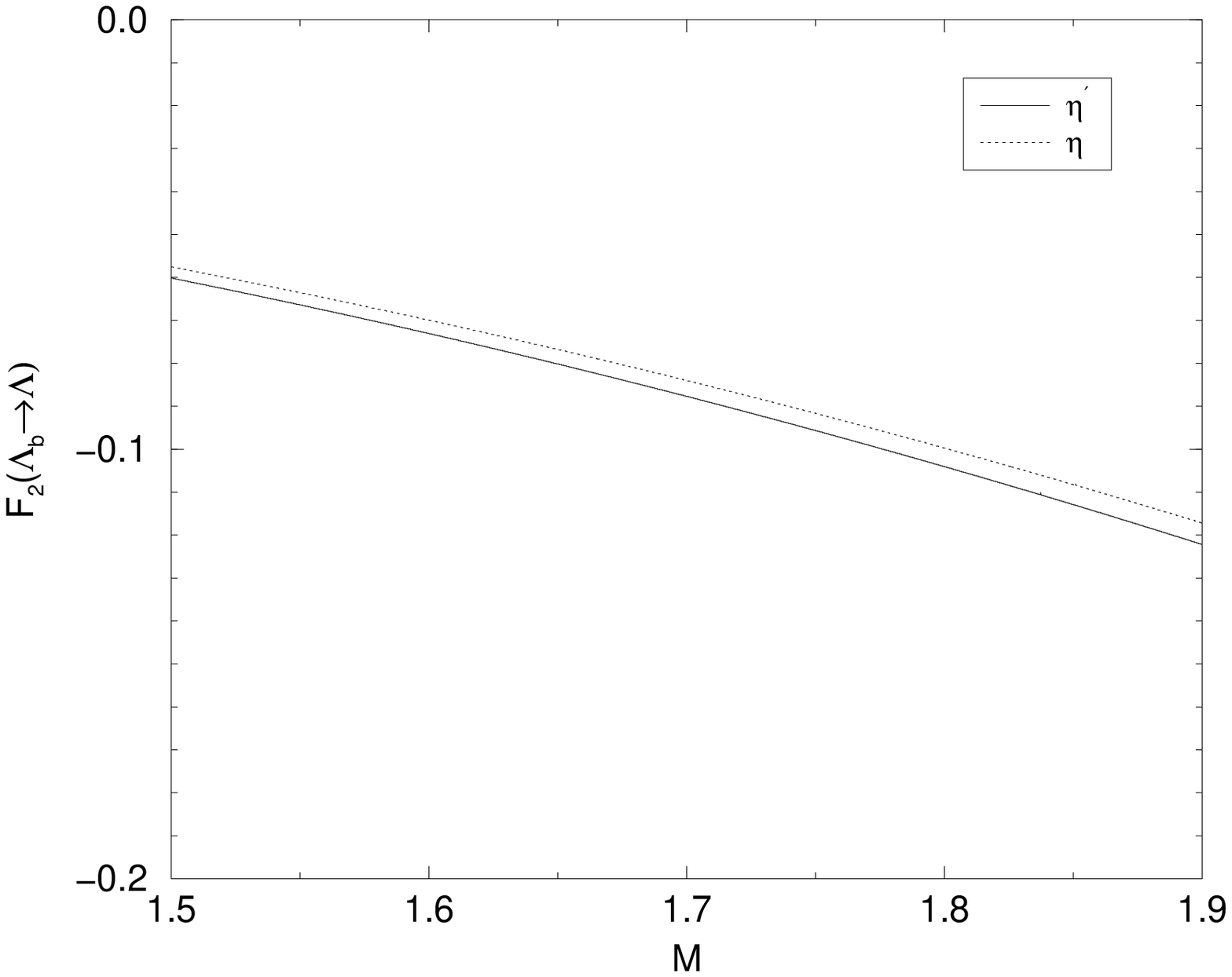 width 10cm)}
    \smallskip
    \caption{\label{fig:fig2} The form factor $F_2$ for the
    transition $\Lambda_b \to \Lambda$ versus the Borel parameter
    $M ~(= {4 T \over m_b})$.
    The dotted (solid) line corresponds to the case of
    $\Lambda_b \to \Lambda \eta^{(')}$.}
\end{figure}
%%%%%%%%%%%%%%%%%%%%%%%%%%%%%%%%%%%%%%%%%%%

We note that the BR of $\Lambda_b \to \Lambda \eta$ is similar to
that of $\Lambda_b \to \Lambda \eta'$, in contrast to the case of
$B \to K \eta^{(')}$ where the BR of $B \to K \eta$ is about an
order of magnitude smaller than that of $B \to K \eta'$.  This
difference mainly arises from the fact that in the factorization
scheme, the decay amplitude for $\Lambda_b \to \Lambda \eta^{(')}$
consists of terms proportional to $\langle \eta^{(')} | O | 0
\rangle \langle \Lambda | O' | \Lambda_b \rangle$ only (see Eq.
(\ref{amp})), while the decay amplitude for $B \to K \eta^{(')}$
consists of terms proportional to $\langle K | \tilde O | 0
\rangle \langle \eta^{(')} | \tilde O' | B \rangle$ as well as
terms proportional to $\langle \eta^{(')} | O | 0 \rangle \langle
K | O' | B \rangle$ \cite{ali2,chen,ddo}. (Here $O^{(')}$ and
$\tilde O^{(')}$ denote the relevant quark currents arising from
the effective Hamiltonian (\ref{effham}).) In the case of $B \to K
\eta^{(')}$, the destructive (constructive) interference appears
between the penguin amplitude proportional to $\langle K | \tilde
O | 0 \rangle \langle \eta^{(')} | \tilde O' | B \rangle$ and that
proportional to $\langle \eta^{(')} | O | 0 \rangle \langle K | O'
| B \rangle$, due to the opposite (same) sign between $\langle K |
\tilde O | 0 \rangle \propto f_K$ and $\langle \eta^{(')} | O | 0
\rangle \propto f^q_{\eta^{(')}}$: in particular, $f^s_{\eta} =
-112$ MeV, while $f^s_{\eta'} = +137$ MeV (see Eq. (\ref{femix})).
However, in the case of $\Lambda_b \to \Lambda \eta^{(')}$, there
is no such interference between terms in the amplitude because the
amplitude contains terms proportional to $\langle \eta^{(')} | O |
0 \rangle \propto f^q_{\eta^{(')}}$ only\footnote{In fact, in
$\Lambda_b \to \Lambda \eta^{(')}$ there is some interference
between the penguin amplitudes proportional to $f^u_{\eta^{(')}}$
and $f^s_{\eta^{(')}}$ (see Eq. (\ref{amp})).  But, it turns out
that the interference does not make a sizable difference between
the BRs of $\Lambda_b \to \Lambda \eta'$ and $\Lambda_b \to
\Lambda \eta$~.}.

%%%%%%%%%%%%%%%%%%%%%%%%%%%%%%%%%%%%%%%%%%%
\begin{figure}
    \centerline{ \DESepsf(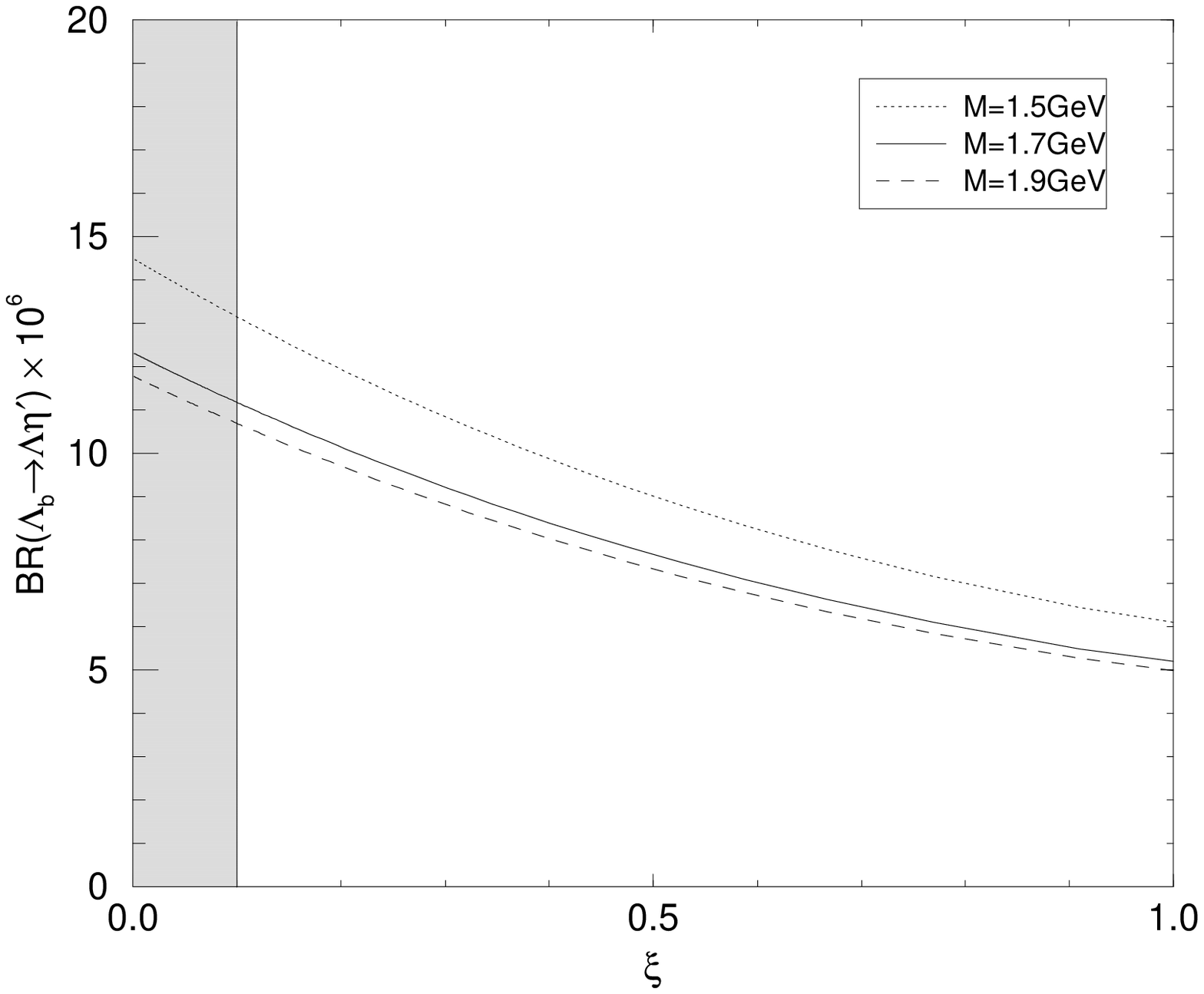 width 10cm)}
    \smallskip
    \caption{\label{fig:fig3} The BR for the decay $\Lambda_b
    \to \Lambda \eta'$ versus $\xi = {1 \over N_c}$ for different values
    of the Borel parameter $M = {4 T \over m_b}$. The shaded region
    denotes the case of $\xi \leq 0.1$, which is favored from the
    analysis of $B \to K \eta'$ decays.}
\end{figure}
%%%%%%%%%%%%%%%%%%%%%%%%%%%%%%%%%%%%%%%%%%%

%%%%%%%%%%%%%%%%%%%%%%%%%%%%%%%%%%%%%%%%%%%
\begin{figure}
    \centerline{ \DESepsf(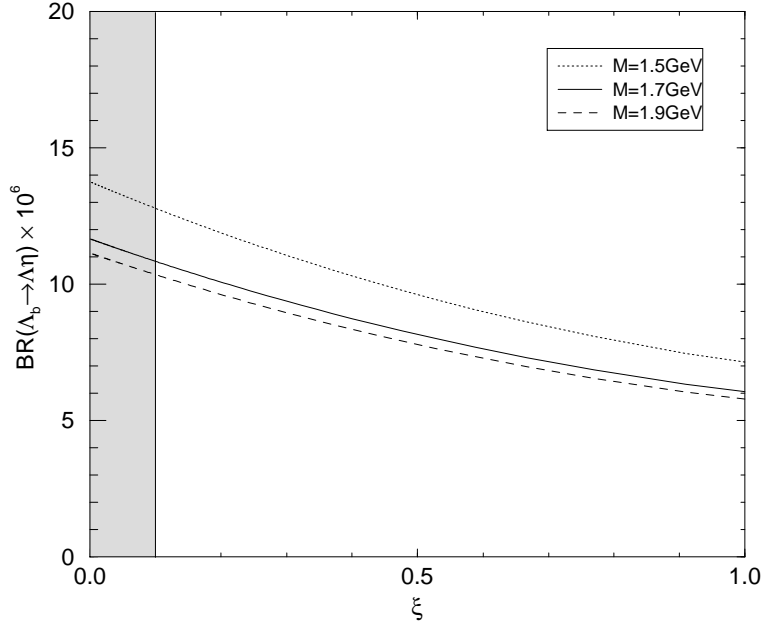 width 10cm)}
    \smallskip
    \caption{\label{fig:fig4} The BR for the decay $\Lambda_b
    \to \Lambda \eta$ versus $\xi = {1 \over N_c}$ for different values
    of the Borel parameter $M = {4 T \over m_b}$. The shaded region
    denotes the case of $\xi \leq 0.1$, which is favored from the
    analysis of $B \to K \eta'$ decays.}
\end{figure}
%%%%%%%%%%%%%%%%%%%%%%%%%%%%%%%%%%%%%%%%%%

In the pole model \cite{mannel,geng}, the form factors are given
by
\begin{equation}
F_i(q^2) = N_i \left( \frac{\Lambda_{\rm QCD}}{\Lambda_{\rm
QCD}+z} \right)^2 ~, ~~ (i = 1, 2)
\end{equation}
where $\Lambda_{\rm QCD}\sim $ 200 MeV and $z= \frac{p_\Lambda
\cdot p_{\Lambda_b}}{m_{\Lambda_b}}$. Using $N_1 = 52.32$ and $N_2
= -13.08$~, we obtain the values of the form factors: $F_1 (q^2) =
0.225 ~(0.217)$ and $F_2 (q^2) = -0.056 ~(-0.054)$ for $q^2 =
m^2_{\eta'} ~(m^2_{\eta})$. We note that the magnitudes of these
form factors are less than a half of those obtained in the QCD sum
rule method. This would result in the fact that the BRs for
$\Lambda_b \to \Lambda \eta^{(')}$ predicted in the case of the
pole model are quite smaller than those predicted in the case of
the QCD sum rule approach.  Indeed, the BRs for $\Lambda_b \to
\Lambda \eta'$ and $\Lambda_b \to \Lambda \eta$ are estimated to
be
\begin{equation}
{\cal B} (\Lambda_b \to \Lambda \eta') = (1.7 - 4.0) \times 10^{-6}~,
\end{equation}
and
\begin{equation}
{\cal B} (\Lambda_b \to \Lambda \eta) = (1.8 - 3.5) \times 10^{-6}~,
\end{equation}
which are about a quarter of those estimated in
the QCD sum rule case. For $\xi = 1 /3$, ${\cal B} (\Lambda_b \to
\Lambda \eta') = 2.56 \times 10^{-6}$ and ${\cal B} (\Lambda_b \to
\Lambda \eta) = 2.36 \times 10^{-6}$. For $\xi = 0.1$, ${\cal B}
(\Lambda_b \to \Lambda \eta') = 3.15 \times 10^{-6}$ and ${\cal B}
(\Lambda_b \to \Lambda \eta) = 2.77 \times 10^{-6}$.

%%%%%%%%%%%%%%%%%%%%%%%%%%%%%%%%%%%%%%%%%%%%%%%%%%%%%%%%%%
%%%%%%%%%%%%%%%%%%%%%%%%%%%%%%%%%%%%%%%%%%%%%%%%%%%%%%%%%%

\section{Nonspectator Contribution to $\Lambda_b \to \Lambda \eta^{\prime}$ Decay}

To simplify the relevant matrix element of the effective
Hamiltonian (\ref{hnonsp}) for the baryonic decay mode
$\Lambda_b\to\Lambda\eta'$, we use an approximation method where
the strong and weak vertices are factorized.  Therefore, the
amplitude of this decay channel, which is depicted in Fig. 5, can
be written as:
\begin{equation}
A(\Lambda_b\to\Lambda\eta')=g_{\Lambda_bNB}A(B\to
K\eta')g_{\Lambda NK}\; , \label{ampnonsp}
\end{equation}
where $g_{\Lambda_bNB}$ and $g_{\Lambda NK}$ parameterize the
strong $\Lambda_b$-Nucleon-$B$ meson and $\Lambda$-Nucleon-$K$ meson
vertices, respectively.  In fact, an estimate of the product
$g_{\Lambda_bNB}g_{\Lambda NK}$ can be obtained by applying the
same approximation method to the experimentally measured
$\Lambda_b\to\Lambda J/\psi$ decay mode where the decay amplitude
has a similar form as Eq. (\ref{ampnonsp}):
\begin{equation}
A(\Lambda_b\to\Lambda J/\psi)=g_{\Lambda_bNB}A(B\to K
J/\psi)g_{\Lambda NK}\; .
\end{equation}
Consequently, the ratio of the decay rates for
$\Lambda_b\to\Lambda J/\psi$ and $B\to K J/\psi$ can be expressed
as
\begin{equation}
\frac{\Gamma (\Lambda_b\to\Lambda J/\psi)}{\Gamma (B\to K
J/\psi)}=(g_{\Lambda_bNB}g_{\Lambda
NK})^2\frac{m_Bg(m_{\Lambda_b},m_\Lambda
,m_{J/\psi})}{m_{\Lambda_b}g(m_B,m_K ,m_{J/\psi})}\; ,
\end{equation}
where $$g(x,y,z)=\left[
\left( 1- \left( \frac{y}{x} \right)^2- \left( \frac{z}{x} \right)^2 \right)^2
-4 \left( \frac{y}{x} \right)^2 \left( \frac{z}{x} \right)^2 \right]^{1/2}\;\;,
$$ and the second factor on the right hand side is the ratio of the phase space
factors for the corresponding
two-body decays. Inserting the experimental values
${\cal B}(\Lambda_b\to J/\psi\Lambda )=4.7\times 10^{-4}$ and ${\cal B}(B\to
J/\psi K )=1.01\times 10^{-3}$ \cite{pdg2002} in the above ratio
leads to the following estimate:
\begin{equation}
(g_{\Lambda_bNB}g_{\Lambda NK})^2\approx 0.55. \label{gg2}
\end{equation}

%%%%%%%%%%%%%%%%%%%%%%%%%%%%%%%%%%%%%%%%%%%
\begin{figure}
    \centerline{ \DESepsf(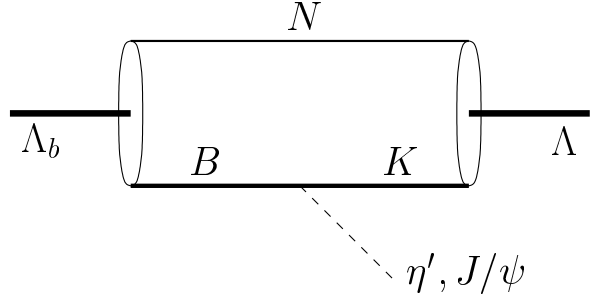 width 8cm)}
    \smallskip
    \caption{\label{fig:fig5} Schematic diagram for the decay $\Lambda_b
\to \Lambda \eta'(J/\psi )$ divided into weak and strong vertices.
}
\end{figure}
%%%%%%%%%%%%%%%%%%%%%%%%%%%%%%%%%%%%%%%%%%

On the other hand, the decay rate for $B\to K\eta'$ via the
nonspectator Hamiltonian (\ref{hnonsp}) can be calculated as
\cite{ahmady2}
\begin{equation}
\Gamma (B\to K\eta')=\frac{C^2H^2f_B^2f_K^2}{384\pi
p^4}\left[\frac{(N_c^2-1)^2}{N_c^4}\right] \vert \vec
p_{\eta'}\vert^3\left[ 3p_\circ^2 \vert \vec
p_{\eta'}\vert^2+(m_{\eta'}^2+\vert \vec
p_{\eta'}\vert^2)(p_\circ^2-p^2)\right]\; , \label{gbketap}
\end{equation}
where $\vert \vec p_{\eta'}\vert$ is the three
momentum of the $\eta'$ meson, i.e.
\begin{equation}
\vert \vec p_{\eta'}\vert =\left[
\frac{(m_B^2+mK^2-m_{\eta'}^2)^2}{4m_B^2}-m_K^2\right]^{1/2}\; ,
\label{petap}
\end{equation}
and $p_\circ$ is the energy transfer by the gluon emitted from the light quark
in the $B$ meson rest frame.  As a result, using Eq.
(\ref{ampnonsp}), one can calculate the ratio of the decay rates
for $\Lambda_b\to\Lambda\eta'$ and $B\to K\eta'$ in terms of the
strong couplings $g_{\Lambda_bNB}$ and $g_{\Lambda NK}$:
\begin{equation}
\frac{\Gamma (\Lambda_b\to\Lambda\eta')}{\Gamma (B\to
K\eta')}=0.91(g_{\Lambda_bNB}g_{\Lambda NK})^2\; .
\label{gratio}
\end{equation}
The numerical factor in Eq. (\ref{gratio}) is due to the phase
space difference as $m_B$ and $m_K$ are replaced by
$m_{\Lambda_b}$ and $m_\Lambda$ in Eq. (\ref{petap}) for the
former decay mode.  In fact, as long as the experimental data (the
average of Eq. (1)) is used to constrain the model parameters
$p_0$ and $p^2$ via Eq. (\ref{gbketap}), the ratio of the rates in
Eq. (\ref{gratio}) turns out to be quite insensitive to these
parameters. As a result, the change in the numerical factor of Eq.
(\ref{gratio}) for the reasonable range of $0.1-0.5$ GeV is less
than $1\%$. At the same time, the approximation which is depicted
in Fig. 5 leads to the cancellation of all the multiplicative
model parameters such as $N_c$.  Inserting Eq. (\ref{gg2}) in Eq.
(\ref{gratio}) and using the input ${\cal B}(B\to K\eta')=(75\pm
8)\times 10^{-6}$, which is obtained from the experimental data
(1), leads to our estimate of the $\eta'$ production in the
$\Lambda_b\to\Lambda$ transition:
\begin{equation}
{\cal B}(\Lambda_b\to\Lambda\eta' )\approx (37.5\pm 4.0)\times
10^{-6}\; .
\end{equation}

%%%%%%%%%%%%%%%%%%%%%%%%%%%%%%%%%%%%%%%%%%%%%%%%%%%%%%%%%%
%%%%%%%%%%%%%%%%%%%%%%%%%%%%%%%%%%%%%%%%%%%%%%%%%%%%%%%%%%

\section{Conclusions}
In this work, we calculated the BRs for the two-body
hadronic decays of $\Lambda_b$ to $\Lambda$ and $\eta$ or $\eta'$
mesons.  The form factors of the relevant hadronic matrix elements
are evaluated by two methods: QCD sum rules and the pole model. In
QCD sum rules, the sensitivity of the form factors to the Borel
parameter is roughly the same for $\eta$ and $\eta'$. The
variation of $F_1$ is around $7\%$ for the Borel parameter in the
range between 1.5 and 1.9.  $F_2$ on the other hand, is quite
sensitive to this parameter, changing by a factor 2 approximately,
in the above range.  Also, we have checked the variation of the
BRs for $\Lambda_b\to\Lambda\eta^{(\prime )}$ with
the effective number of colors $N_c$ in order to extend our results
to $\xi =\frac{1}{N_c}\leq 0.1$ range, which is favored in fitting the
experimental data on the ${\cal B}(B\to K\eta' )$ in the framework
of generalized factorization.  Our results indicate that the
BRs for $\Lambda_b\to\Lambda\eta$ and
$\Lambda_b\to\Lambda\eta'$ are more or less the same in QCD sum
rules, $9.15 \times 10^{-6}$ and $8.93 \times 10^{-6}$,
respectively, for $M=1.7$ GeV and $N_c=3$.

In the pole model on the other hand, the form factor $F_1$ turns
out to be smaller approximately by a factor 2.  However, $F_2$ is
roughly the same as in the sum rule case for the smaller values of
the Borel parameter.  As a result, the predicted branching ratios
in this model, ${\cal B}(\Lambda_b\to \Lambda\eta ) =2.36 \times
10^{-6}$ and ${\cal B}(\Lambda_b\to \Lambda\eta' )=2.56 \times
10^{-6}$ for $N_c=3$, are significantly smaller than those
obtained via QCD sum rules.

We also made an estimate of the nonspectator gluon fusion
mechanism to the hadronic $\Lambda_b\to\Lambda\eta'$ decay.  The
purpose is to find the enhancement of the BR of this
baryonic decay if the same underlying process that leads to an
unexpectedly large BR for $B\to K\eta'$ is operative
in this case as well.  We used a simple approach for this estimate
where the amplitude is divided into strong and weak vertices.  Our
results point to a substantial increase in the BR,
from more than a factor 3 to around an order of magnitude,
compared to QCD sum rule and the pole model predictions,
respectively.  Future measurements of this $\Lambda_b$ decay mode
will test the extent of the validity of these models.

%%%%%%%%%%%%%%%%%%%%%%%%%%%%%%%%%%%%%%%%%%%%%%%%%%%%%%
%%%%%%%%%%%%%%%%%%%%%%%%%%%%%%%%%%%%%%%%%%%%%%%%%%%%%%
\vspace{1cm}
\centerline{\bf ACKNOWLEDGEMENTS}
\medskip

\noindent The work of C.S.K. was supported
by Grant No. 2001-042-D00022 of the KRF.
The work of S.O. was supported in part by Grant No. R02-2002-000-00168-0
from BRP of the KOSEF, and by the Japan Society for the
Promotion of Science (JSPS).
The work of C.Y. was supported
in part by  CHEP-SRC Program,
in part by Grant No. R02-2002-000-00168-0 from BRP of the KOSEF.
M.R.A. is grateful to the Natural Sciences \& Engineering Research Council of
Canada (NSERC) for financial support.

\newpage

%%%%%%%%%%%%%%%%%%%%%%%%%%%%%%%%%%%%%%%%%%%%%%%%%%%%%%%%%%
%%%%%%%%%%%%%%%%%%%%%%%%%%%%%%%%%%%%%%%%%%%%%%%%%%%%%%%%%%


\begin{thebibliography}{99}
\bibitem{cleo} S. J. Richichi \etal ~(CLEO Collaboration), Phys. Rev. Lett. {\bf
85}, 520 (2000).
%%CITATION = HEP-EX 9912059;%%

\bibitem{belle} K. F. Chen (Belle Collaboration), talk at the
31$^{\rm th}$ International Conference on High Energy Physics,
Amsterdam, Netherlands, July, 2002.

\bibitem{babar} A. Bevan (BaBar Collaboration), talk at the
31$^{\rm th}$ International Conference on High Energy Physics,
Amsterdam, Netherlands, July, 2002.

\bibitem{kagan} A. L. Kagan and A. A. Petrov, UCHEP-27, UMHEP-443, hep-ph/9707354;
%%CITATION = HEP-PH 9707354;%%
A. Datta, X.-G. He and S. Pakvasa, Phys. Lett. B {\bf 419}, 369
(1998).
%%CITATION = HEP-PH 9707259;%%

\bibitem{atwood} D. Atwood and A. Soni, Phys. Lett. B {\bf 405}, 150
(1997);
%%CITATION = HEP-PH 9704357;%%
W. -S. Hou and B. Tseng, Phys. Rev. Lett. {\bf 80} 434 (1998).
%%CITATION = HEP-PH 9705304;%%

\bibitem{ali1} A. Ali, J. Chay, C. Greub and P. Ko, Phys. Lett. B {\bf 424}, 161
(1998).
%%CITATION = HEP-PH 9712372;%%

\bibitem{halperin} I. Halperin and A. Zhitnitsky, Phys. Rev. Lett. {\bf 80} 438 (1998).
%%CITATION = HEP-PH 9705251;%%

\bibitem{ali2} A. Ali and C. Greub, Phys. Rev. D {\bf 57}, 2996 (1998)
and references therein.
%%CITATION = HEP-PH 9707251;%%

\bibitem{cheng1} H.-Y. Cheng and B. Tseng, Phys. Lett. B {\bf 415}, 263 (1997).
%%CITATION = HEP-PH 9707316;%%

\bibitem{du} D. Du, C. S. Kim and Y. Yang, Phys. Lett. B {\bf 426}, 133
(1998).
%%CITATION = HEP-PH 9711428;%%

\bibitem{ahmady1} M. R. Ahmady, E. Kou and A. Sugamoto, Phys. Rev. D {\bf 58},
014015 (1998).
%%CITATION = HEP-PH 9710509;%%

\bibitem{kou} E. Kou and A. I. Sanda, Phys. Lett. B {\bf 525}, 240 (2002).
%%CITATION = HEP-PH 0106159;%%

\bibitem{beneke} M. Beneke and M. Neubert, Nucl. Phys. B {\bf 651}, 225 (2003).
%%CITATION = HEP-PH 0210085;%%

\bibitem{koy} For determination of the flavor-singlet contribution, as proposed in
Ref.~\cite{beneke}, whose unknown
value prevents accurate theoretical estimates in analysis of $B \to \eta' K$ decays
in QCD factorization, please look at \\
C. S. Kim, Sechul Oh and Chaehyun Yu, [hep-ph/0305032].
%%CITATION = HEP-PH 0305032;%%

\bibitem{choudhury} D. Choudhury, B. Dutta and Anirban Kundu, Phys. Lett. B {\bf 456},
185 (1999).
%%CITATION = HEP-PH 9812209;%%

\bibitem{dko1} B. Dutta, C. S. Kim and Sechul Oh, Phys. Lett. B {\bf 535}, 249
(2002).
%%CITATION = HEP-PH 0202019;%%

\bibitem{dko2} B. Dutta, C. S. Kim and Sechul Oh, Phys. Rev. Lett. {\bf 90}, 011801
(2003).
%%CITATION = HEP-PH 0208226;%%

\bibitem{stone} S. Stone, in Proceedings of the Third
International Conference on B Decays and CP Violation, edited by
H.-Y. Cheng and W.-S. Hou (World Scientific, 2000), p. 450,
[hep-ph/0002025].
%%CITATION = HEP-PH 0002025;%%

\bibitem{giri} A. K. Giri, R. Mohanta and M. P. Khanna, Phys.
Rev. D {\bf 65}, 073029 (2002).
%%CITATION = HEP-PH 0112220;%%

\bibitem{aod} D. Buskulic \etal (ALEPH Collaboration), Phys. Lett.
B {\bf 365}, 437 (1996);
%%CITATION = PHLTA,B365,437;%%
G. Abbiendi \etal (OPAL Collaboration), Phys. Lett. B {\bf 444},
539 (1998);
%%CITATION = HEP-EX 9808006;%%
P. Abreu \etal (DELPI Collaboration), Phys. Lett. B {\bf 474}, 205
(2000).
%%CITATION = PHLTA,B474,205;%%

\bibitem{chen} Y.-H. Chen, H.-Y. Cheng, B. Tseng and K.-C. Yang, Phys. Rev. D
{\bf 60}, 094014 (1999).
%%CITATION = HEP-PH 9903453;%%

\bibitem{hou} W. S. Hou, Nucl. Phys. B {\bf 308}, 561 (1988).
%%CITATION = NUPHA,B308,561;%%

\bibitem{ddo} N. G. Deshpande, B. Dutta and Sechul Oh, Phys. Rev. D {\bf
57}, 5723 (1998);
%%CITATION = HEP-PH 9710354;%%
N. G. Deshpande, B. Dutta and Sechul Oh, Phys. Lett. B {\bf 473},
141 (2000).
%%CITATION = HEP-PH 9712445;%%

\bibitem{bensalem} W. Bensalem, A. Datta and D. London, Phys.
Lett. B {\bf 538}, 309 (2002).
%%CITATION = HEP-PH 0205009;%%

\bibitem{ahmady2} M. R. Ahmady and E. Kou, Phys. Rev. D {\bf 59},
054014 (1999).
%%CITATION = HEP-PH 9807398;%%

\bibitem{leut} H. Leutwyler, Nucl. Phys. B (Proc. Suppl.) {\bf 64}, 223
(1998).
%%CITATION = HEP-PH 9709408;%%

\bibitem{feldmann} T. Feldmann, P. Kroll and B. Stech, Phys. Rev.
D {\bf 58}, 114006 (1998);
%%CITATION = HEP-PH 9802409;%%
Phys. Lett. B {\bf 449}, 339 (1999).
%%CITATION = HEP-PH 9812269;%%

\bibitem{ball} P. Ball, J. M. Frere and M. Tytgat, Phys. Lett. B
{\bf 365}, 367 (1996).
%%CITATION = HEP-PH 9508359;%%

\bibitem{cheng2} H.-Y. Cheng and B. Tseng, Phys. Rev. D {\bf 58}, 094005 (1998).
%%CITATION = HEP-PH 9803457;%%

\bibitem{huang} C.-S. Huang and H.-G. Yan, Phys. Rev. D {\bf 59},
114022 (1999); {\it Erratum-ibid}, D {\bf 61}, 039901 (2000).
%%CITATION = HEP-PH 9811303;%%

\bibitem{mannel} T. Mannel, W. Roberts and Z. Ryzak, Nucl. Phys. B {\bf 355}, 38
(1991).
%%CITATION = NUPHA,B355,38;%%

\bibitem{geng} C.-H. Chen and C. Q. Geng, Phys. Lett. B {\bf 516},
327 (2001).
%%CITATION = HEP-PH 0101201;%%

\bibitem{pdg2002} K. Hagiwara \etal ~(Particle Data Group), Phys.
Rev. D {\bf 66}, 010001 (2002).
\end{thebibliography}
\end{document}